\documentstyle[aps,prb,multicol,epsf,graphics]{revtex}

\newcommand \be {\begin{equation}}
\newcommand \ee {\end{equation}}

\newcommand \bi {\bibitem}
\newcommand \s {\sigma}

\begin{document}

\title{A conjectured scenario for order-parameter fluctuations in spin glasses}

\author{Felix Ritort and Marta Sales}
\address{Department of Physics, Faculty of Physics, University of
Barcelona\\
Diagonal 647, 08028 Barcelona (Spain)
}

\date{\today}

\maketitle

\begin{abstract}
We study order-parameter fluctuations (OPF) in disordered systems by
considering the behavior of some recently introduced paramaters
$G,G_c$ which have proven very useful to locate phase transitions.  We
prove that both parameters $G$ (for disconnected overlap disorder
averages) and $G_c$ (for connected disorder averages) take the
respective universal values $1/3$ and $13/31$ in the $T\to 0$ limit
for any {\em finite} volume provided the ground state is {\em unique}
and there is no gap in the ground state local-field distributions,
conditions which are met in generic spin-glass models with continuous
couplings and no gap at zero coupling. This makes $G,G_c$ ideal
parameters to locate phase transitions in disordered systems much alike
the Binder cumulant is for ordered systems.  We check our results by
exactly computing OPF in a simple example of uncoupled spins in the
presence of random fields and the one-dimensional Ising spin glass. At
finite temperatures, we discuss in which conditions the value 1/3 for
$G$ may be recovered by conjecturing different scenarios depending on
whether OPF are finite or vanish in the infinite-volume limit. In
particular, we discuss replica equivalence and its natural consequence
$\lim_{V\to\infty}G(V,T)=1/3$ when OPF are finite.  As an example of a
model where OPF vanish and replica equivalence does not give
information about $G$ we study the Sherrington-Kirkpatrick spherical
spin-glass model by doing numerical simulations for small
sizes. Again we find results compatible with $G=1/3$ in the spin-glass
phase.
\end{abstract} 


%\begin{multicols}{2}
%\narrowtext

\section{INTRODUCTION}

It is well known that mean-field spin glasses are characterized by
strong (non vanishing in the thermodynamic limit) sample to sample
fluctuations of the order parameter \cite{BOOKS}. Despite the fact
that extensive thermodynamic quantities (such as free energy and all
its finite order derivatives) are self-averaging in the thermodynamic
limit (i.e. their intensive part does not depend on the realization of
the quenched randomness) the same result cannot be extended to order
parameter fluctuations. It is widely believed that absence of
self-averageness of the order parameter is strongly related to replica
symmetry breaking, i.e. the existence of several ergodic components
not related by any symmetry of the Hamiltonian.

Recently, Guerra suggested \cite{GUERRA1} that sample to sample
fluctuations of the order parameter (hereafter referred to as OPF)
verify some sum rules which are generally valid in any disordered
system. This claim assumes that the system is stochastically stable in
the presence of a mean-field perturbation, a property which may
strongly depend on the nature of the equilibrium state. A system is
stochastically stable if its properties (static or dynamic) smoothly
change in the presence of a small random perturbation.  These sum
rules have been recently used to define a new dimensionless parameter
(hereafter called $G$) which takes into account sample to sample
fluctuations \cite{MNPPRZ}. This parameter has been shown to provide
an alternative and powerful way to locate phase transition points in
disordered systems. The advantage of $G$ respect to more canonical
ones (such as the Binder cumulant ratio used in ordered systems)
relies on the fact that it works very well also in the absence of
time-reversal symmetry in the Hamiltonian or other more complex
disordered systems. In particular, the method has been recently
applied for Ising spin glasses \cite{MNPPRZ,PPR}, Migdal-Kadanoff spin
glasses \cite{BBDM}, Potts glasses \cite{POTTS}, Heisenberg spin
glasses, which display a chiral phase transition \cite{HK} as well as
some protein folding models \cite{PPRICCI}.

The purpose of this paper is to show, by using general arguments,
analytic computations for simple models and numerical simulations,
that indeed this new parameter is the appropiate tool to investigate
phase transitions in disordered systems much like the Binder cumulant
is for ordered systems. We conjecture and prove that this parameter
$G$ takes the universal value 1/3 at zero temperature for any
disordered system (finite or infinite) with the only condition of {\em
uniqueness} of the ground state and absence of a zero-temperature gap
in the local field distribution. This condition is satisfied by all
spin-glass models with continuous distribution of couplings and no gap
at zero coupling. At finite temperature $G$ certainly depends on the
system size. We claim that due to the property of replica equivalence,
for models in which OPF are finite, $G$ converges in the
infinite-volume limit to zero if the system is in a paramagnetic phase
and to the same zero-temperature value 1/3 if the system is in the
spin-glass phase. When OPF vanish this does not necessarily hold and
we discuss in what conditions the universal value 1/3 may be
recovered.

%This is a quite strong statement which will be better
%precised in forthcoming sections and its usefulness will be more lengthy
%discussed.

The paper is organized as follows. Section II is a reminder of the
definition of the $G$ parameter as well as some other useful ones.
Section III presents a detailed computation on a simple disordered
model which serves as an illustrative example of the main
results. Section IV proves the zero-temperature conjecture under some
general conditions for any disordered system. Section V presents
detailed calculations on the one-dimensional Ising spin-glass model
using the transfer matrix approach. Section VI addresses the validity
of the conjecture at finite temperature by studying the
Sherrington-Kirkpatrick spherical spin glass, a model where OPF
vanish. Finally we discuss the results and present the conclusions.

\section{The G parameter and replica equivalence}
 
The definition of the $G$ parameter is based on some exact relations
obtained for the sample to sample fluctuations of the order parameter in
the Sherrington-Kirkpatrick (SK) model \cite{BOOKS}. The SK model is
defined by the disordered mean-field Hamiltonian,

\be
{\cal H}_{SK}=-\sum_{i<j}\,J_{ij}\,\s_i\,\s_j~~~~~~,
\label{HSK}
\ee

where the $J_{ij}$ are quenched Gaussian variables with zero
average and variance $1/N$ where $N$ is the number of sites. The SK
model presents a second order phase transition at $T_c=1$ below which
replica symmetry breaks down and ergodicity is broken. The spin-glass phase
is described by an order parameter function $P_J(q_{12})$ where
$q_{12}=\sum_{i=1}^N\s^1_i\s^2_i$ is the replica overlap and the
subindex $J$ stands for the realization of the quenched
randomness. $P_J(q)$ is a simple object in the paramagnetic phase above
$T_c$ ($P_J(q)=\delta(q)$) but develops strong sample to sample
fluctuations below $T_c$ inside the spin-glass phase. Fluctuations in
the order parameter were originally derived by Bray, Moore and Young
\cite{BMY} using the property of replica equivalence
\cite{PARISI1}. This property states that the sum of all elements
contained in a given row (or column) in the replica matrix $Q_{ab}$ is
independent of the row (or column). As shown by Parisi \cite{PARISI1}
this is a necessary condition for the replicated free energy to be
proportional to the number of replicas $n$ and have a well defined free
energy in the limit $n\to 0$. Fluctuations are then described by the
following exact relation in the $N\to\infty$ limit \cite{MPSTV},

\be
\overline{P_J(q_{12},q_{34})}=\frac{1}{3}\overline{P_J(q_{12})}
\delta(q_{12}-q_{34})+\frac{2}{3}\overline{P_J(q_{12})}\overline{P_J(q_{34})}~~~,
\label{eq2}
\ee

where $\overline{(.)}$ stands for disorder average and $1,2,3,4$ denote
replica indices. Therefore,

\be
\overline{P_J(q_{12},q_{34})}\neq\overline{P_J(q_{12})}\overline{P_J(q_{34})}~~~,
\label{eq3}
\ee

so $P_J$ fluctuates with $J_{ij}$ in a non-trivial way. Multiplying both
sides of eq.(\ref{eq2}) by $q_{12}^2$ and $q_{34}^2$ and integrating
over all possible values of the overlaps $q_{12},q_{34}$ one obtains the
following sum rule \cite{BMY}, 

\be
\overline{\langle q_{12}^2\rangle^2}=\frac{1}{3}
\overline{\langle q_{12}^4\rangle}+\frac{2}{3}\overline{\langle
q_{12}\rangle}^2~~~~~~.
\label{eq4}
\ee

where $\langle...\rangle$ stands for thermal average. This
relationship has been also rederived by Guerra using general arguments
based on self-averaging properties of the internal energy as well as
its finite derivatives \cite{GUERRA1}. Now let us define the following
ratio,

\be
G=\frac{\overline{\langle q^2\rangle^2}-\overline{\langle q^2\rangle}^2}
{\overline{\langle q^4\rangle}-\overline{\langle q^2\rangle}^2}
\label{eq5}
\ee

%Due to the sum rule (\ref{eq5}) the new parameter defined can take only
%two values. 

Note that the numerator in (\ref{eq5}) corresponds (except for the
absence of a multiplicative constant $N^2$) to the sample fluctuations
of the spin-glass susceptibility.  For the SK model, because of the sum
rule (\ref{eq5}), it is possible to show that $G$ takes only two
values. $G$ is equal to 1/3 in the replica symmetry broken phase and
vanishes above $T_c$,

\be
G=\frac{1}{3}\Theta(T_c-T)~~~~~~~.
\label{eqG}
\ee

The generality of the replica-equivalence property suggests that
(\ref{eqG}) will hold in any system (even beyond mean-field) if OPF do
not vanish in the limit $V\to\infty$.  But may well happen that OPF
vanish. Then both numerator and denominator in (\ref{eq5}) vanish in
the $V\to\infty$ limit. In this case replica equivalence is not enough
to decide what the value of $G$ is. The value of $G$ is then
determined by the form of the finite-size corrections to the order
parameter (and in particular its prefactors), which in principle could
not satisfy sum rules such as (\ref{eq4}).  Despite this uncertainty,
in this paper we propose three possible scenarios for the parameter
$G$,

\begin{itemize}
\item{1.} OPF remain finite in the thermodynamic limit. This is the general
situation encountered in mean-field models with a replica broken phase.
So both numerator and denominator in (\ref{eq5}) are finite in the
infinite volume limit. The property of replica equivalence and also
stochastic stability indicate that the same should be valid for any
finite-dimensional disordered system (assuming that for those systems
OPF are finite) leading to $G=1/3$ in the spin-glass phase.

\item{2.} OPF vanish in the large volume limit like $1/V$. This is the
situation typically encountered in the paramagnetic phase. The ratio may
then be zero or finite depending on the particular case.

%but the numerator in
%(\ref{eq5}) vanishes faster (as the volume increases) than the
%denominator does. 

\item{3.} OPF vanish in the large volume limit slower than $1/V$ (for
instance, like $\frac{1}{V^{\alpha}}$ with $\alpha<1$). This situation
is typical of disordered systems with a marginally stable replica
symmetric phase. Both numerator and denominator in (\ref{eq5}) vanish,
the ratio $G$ is finite but may be different from 1/3 at finite
temperature. In this case the property of replica equivalence cannot be
used for the reason discussed before and stochastic stability may not
hold. Actually the property of stochastic stability may breakdown if the
equilibrium phase is drastically changed in the presence of a mean-field
perturbation.  This situation may be found in spin-glass models without
OPF such as hierarchichal lattices (i.e. spin glasses in the
Migdal-Kadanoff approximation), the Sherrington-Kirkpatrick spherical
spin glass (see section VI) or finite-dimensional models described by a
unique low-temperature state such as the droplet model.

\end{itemize}

%Which possibility among those three is found relies on the particular
%model considered and the stochastic stability property. 
Despite the
main hypothesis of stochastic stability remains still to be proven
all previous three cases seem quite reasonable and we do not know of
non-trivial counterexamples. Note that  there is not any direct
relationship between OPF and the value of $G$ in the low-temperature
phase. Actually, the previous possibilities 1 and 3 may yield the same
value of $G$ although the physical description of the low-temperature
phase is much different. As has been observed in \cite{BBDM} the non
vanishing of $G$ should not be taken as direct evidence for
non-vanishing OPF or replica symmetry breaking. In order to better
evidenciate whether OPF survive in the infinite-volume limit, it is
necessary to consider another dimensionless parameter, which has not the
ambiguity of the ratio of two vanishing quantities.  For instance one
may define the $A$ parameter \cite{MNPPRZ},

\be
A=\frac{\overline{\langle q^2\rangle^2}-\overline{\langle q^2\rangle}^2}
{\overline{\langle q^2\rangle}^2}~~~~~~,
\label{eq6}
\ee

which is nothing else than the numerator of (\ref{eq5}) appropriately
normalized. We will see later that the nice properties of $G$ are not
present in the parameter $A$ and the former is much more convenient
to locate phase transitions. Generally, one expects $A$ to be a non
trivial function of both volume and temperature vanishing (in the
$V\to\infty$ limit) only when OPF vanish (for instance, in a
paramagnetic phase). If OPF are finite $A$ may take a finite value
but an identity such as (\ref{eqG}) for $A$ does not hold.

In this paper we will show examples for all three behaviors, by
explicit analytic computations and some numerical
calculations. Furthermore, we will show that for models with a unique
ground state and without gap in the ground-state local field
distribution,

\be
\lim_{T\to 0}G(V,T)=\frac{1}{3}~~~~~~,
\label{eq7}
\ee

so the $G$ parameter is 1/3 at $T=0$ for {\em any} finite volume
$V$. This is not anymore true at finite temperature where the parameter
$G$ may take the value $1/3$ only in the infinite volume limit.

Before finishing this section let us remind that in references \cite{MNPPRZ,PPR}
another quantities similar to (\ref{eq5}) and (\ref{eq6}) have been
proposed for systems without time-reversal symmetry. These are defined by
considering the {\em connected} overlaps,

\be
G_c=\frac{\overline{\langle (q-\langle q\rangle)^2\rangle^2}-\overline{\langle (q-\langle q\rangle)^2\rangle}^2}
{\overline{\langle (q-\langle q\rangle)^4\rangle}-\overline{\langle (q-\langle q\rangle)^2\rangle}^2}~~~~,
\label{eq8}
\ee

\be
A_c=\frac{\overline{\langle (q-\langle q\rangle)^2\rangle^2}-\overline{\langle (q-\langle q\rangle)^2\rangle}^2}
{\overline{\langle (q-\langle q\rangle)^2\rangle}^2}~~~~~~.
\label{eq9}
\ee

We will see that a result like (\ref{eq7}) applies also to the
parameter $G_c$ and our result reads:

\be
\lim_{T\to 0}G_c(V,T)=\frac{13}{31}~~~~~.
\label{eq9b}
\ee

For the SK model the quantity $G_c$ is defined by restricting thermal
averages to the $q> 0$ part of the $P(q)$. $G_c$ does not satisfy the
identity (\ref{eqG}) so this is not the best quantity to look at in
numerical simulations despite the fact that both $G_c$ and $G$ (and also
$A_c$ and $A$) may take similar values in the vicinity of the critical
region. This explains why similar results were obtained for both sets of
quantities in numerical simulations.

\section{An instructive example}

Here we analyze in detail a solvable case which will be useful to
illustrate the main contents of the paper and how disorder expectation
values of the overlaps are computed. Moreover, the analysis of this
section will prove to be useful for a constructive proof of the zero-temperature
results (\ref{eq7}) and (\ref{eq9b}) to be presented later on. Consider the
following Hamiltonian,

\be
{\cal H}=-\sum_{i=1}^V\,h_i\s_i~~~~~~,
\label{eq10}
\ee

where the spins may take the values $\pm 1$ and the fields $h_i$ are
uncorrelated and randomly taken from a distribution $P(h)$ with
finite weight at zero field (i.e. $P(0)$ finite). In principle $P(h)$
may be any function

\be
P(h)=\tilde{P}(h)+\sum_k c_k \delta(h-h_k)
\label{eq11}
\ee

with $\tilde{P}(h)$ any continuous function, all $h_k\ne 0$ and
 $\tilde{P}(0)\ne 0$. This condition is enough to ensure the
 non-degeneracy of the ground state because there is a single
 configuration which minimizes the energy $\s_i^*={\rm sign}(h_i)$. Note
 that if a finite fraction of the fields $h_i$ were zero then the ground
 state would be degenerate. With this very general condition we may
 exactly compute the parameters $G$ and $A$ introduced in the previous
 section. Everything reduces to compute the three overlap quantities:
 $\overline{\langle q^2\rangle},\overline{\langle q^4\rangle}$ and
 $\overline{\langle q^2\rangle^2}$. The computations are quite
 elementary and here we present the final results. For the numerator and
 denominator of eq.(\ref{eq5}) we get,

\be
Numerator\equiv\frac{2(V-1)}{V^3}\overline{R^2}^2+
\frac{4(V-1)(V-2)}{V^3}\overline{R^2}\,\overline{R}^2-
\frac{2(2V^2-5V+3)}{V^3}\overline{R}^4~~~~~~,
\label{eq12}
\ee
\be
Denominator\equiv \frac{2}{V^2}-\frac{2}{V^3}+
\frac{4(V-1)(V-2)}{V^3}
\overline{R}^2-\frac{2(2V^2-5V+3)}{V^3}\overline{R}^4~~~~,\label{eq13}
\ee

where 

\be
\overline{R}=\int_{-\infty}^{\infty}dhP(h)\tanh^2(\beta h)~~~,\label{eq14}
\ee
\be
\overline{R^2}=\int_{-\infty}^{\infty}dhP(h)\tanh^4(\beta h)~~~,\label{eq15}
\ee

and $P(h)$ is the generic distribution (\ref{eq11}). 
The expressions for the parameters $G$ and $A$ may be further simplified
yielding,

\be
G=\frac{(\overline{R^2}-\overline{R}^2)(\overline{R^2}+(2V-3)\overline{R}^2)}
{(1-\overline{R}^2)(1+(2V-3)\overline{R}^2)}~~~~~,
\label{eq18}
\ee

and

\be
A=\frac{2(V-1)(\overline{R^2}-\overline{R}^2)(\overline{R^2}+(2V-3)\overline{R}^2)}
{V(1+(V-1)\overline{R}^2)^2}~~~.
\label{eq19}
\ee

\noindent
Note that in the limit $V\to\infty$ both numerator and denominators in
(\ref{eq12}) and (\ref{eq13}) vanish. The quantity $A$ also vanishes
like $1/V$ but the ratio $G$ stays finite,

\be
\lim_{V\to\infty} G(V,T)=\frac{\overline{R^2}-\overline{R}^2}{1-\overline{R}^2}~~~.
\label{eq20}
\ee

The finite volume quantity $G(T,V)$ in (\ref{eq18}) satisfies the
conjecture (\ref{eq7}). A simple integration by parts reveals that the
asymptotic low-temperature behavior of $\overline{R}$ and
$\overline{R^2}$ is given by,

\be
\overline{R}=1-TD+O(T^2)~~~,~~~\overline{R^2}=\overline{R}-\frac{T}{3}D+O(T^2)~~~,
\label{eq16}
\ee

where $D$ is a positive constant given by

\be
D=2P(0)~~~~~.
\label{eq17}
\ee

\noindent
Substituting the asymptotic behavior (\ref{eq16}) in (\ref{eq18}) we
obtain $G(V,T=0)=1/3$. Note that the same result is obtained
substituting (\ref{eq16}) in (\ref{eq20}) because in this simple example
the two limits $T\to 0$ and $V\to\infty$ may be interchanged. This is
not generally true: in particular when a phase transition takes place at
$T=0$ the two limits may not be interchanged anymore.

For the parameters $G_c$ and $A_c$ introduce in  (\ref{eq8})(\ref{eq9}) we get:
\be
G_c=\frac{\overline{R^4}-(\overline{R^2})^2}{2 V-2-4(V-2)\overline{R^2}+(2 V-3)(\overline{R^2})^2
-3\overline{R^4}}
\ee
\be
A_c=\frac{\overline{R^4}-(\overline{R^2})^2}{V(1-\overline{R^2})^2}
\ee

\begin{figure}
\begin{center}
\rotatebox{270}
{\epsfxsize=8cm 
\epsffile{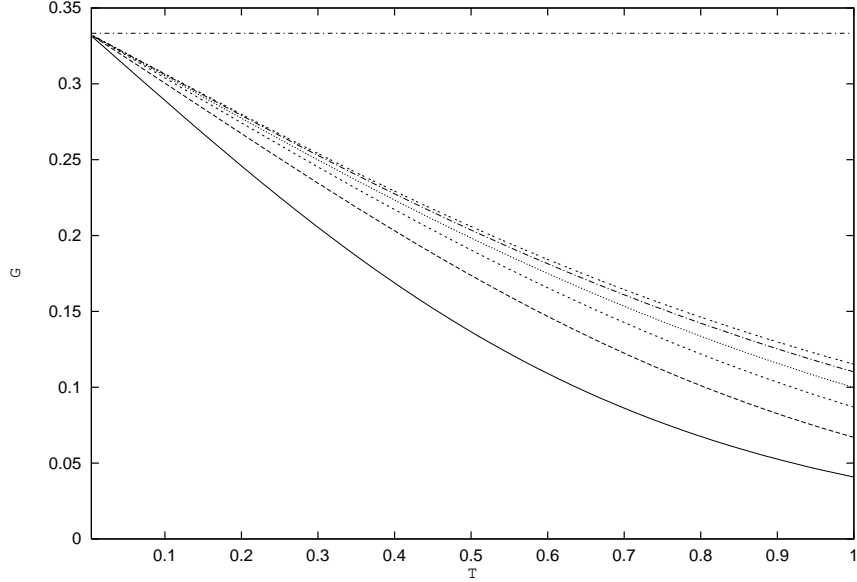}}
\caption{Parameter $G$  for $V=2,4,8,16,50,100$ from below to above}\label{G_example}
\end{center}

\end{figure}

\begin{figure}
\begin{center}
\rotatebox{270}
{\epsfxsize=8cm\epsffile{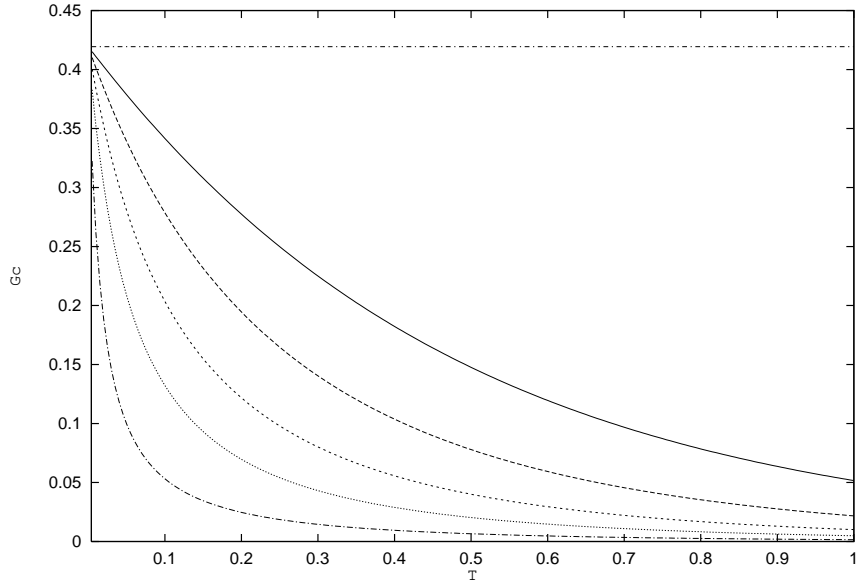}}
\caption{Parameter $G_c$  for $V=2,4,8,16,50$ from below to above}\label{Gc_example}
\end{center}

\end{figure}

\begin{figure}
\begin{center}
\rotatebox{270}
{\epsfxsize=8cm\epsffile{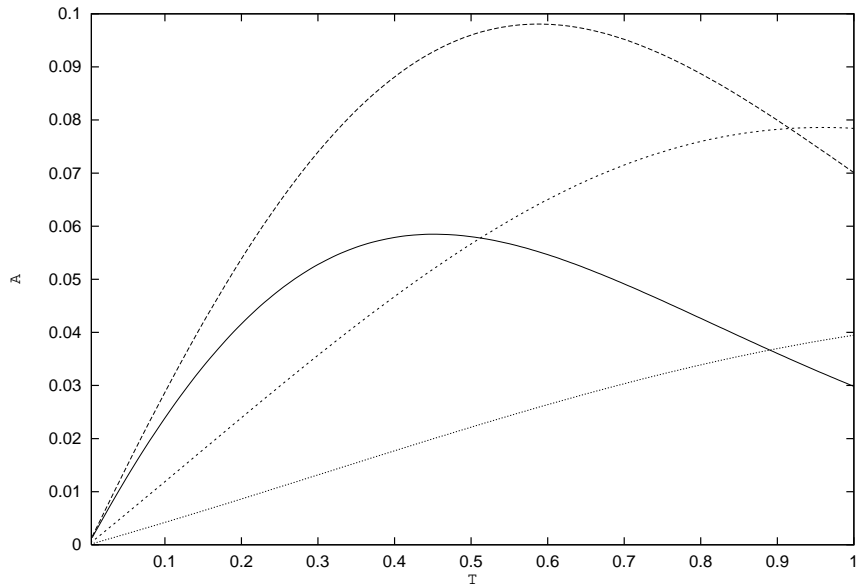}}
\caption{Parameter $A$  for $V=2,4,8,16,50$ from up to down}\label{A_example}
\end{center}

\end{figure}

\begin{figure}
\begin{center}
\rotatebox{270}
{\epsfxsize=8cm\epsffile{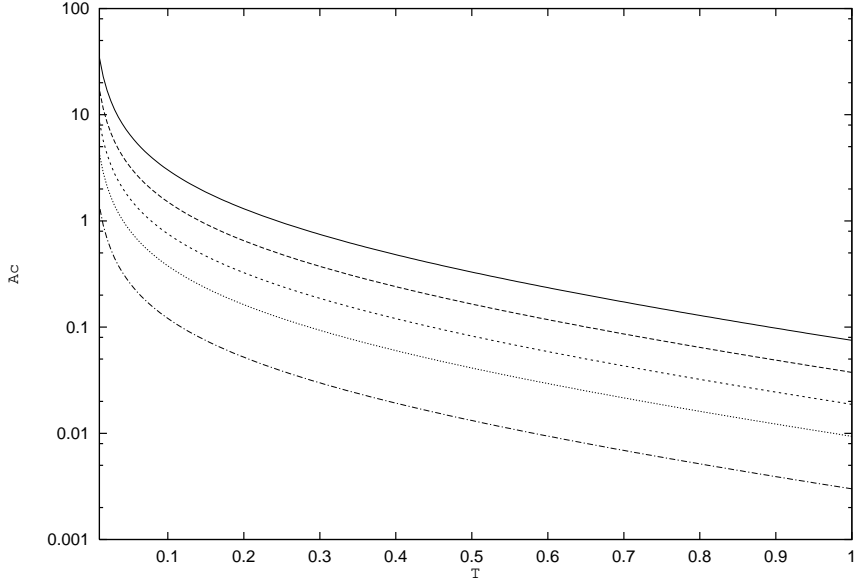}}
\caption{Parameter $A_c$  for $V=2$ (continuous), 4 (long-dashed), 16 (short-dashed), 
50 (dotted)}\label{Ac_example}
\end{center}

\end{figure}

We observe that $G_c$  behaves in a different
way. It tends to zero for $T$ finite and is again
independent of the volume for $T=0$ but takes the value $13/31$. For
$G_c$ the two limits ($V\to\infty$ and $T\to 0$) now do not commute.
Figures~\ref{G_example},~\ref{Gc_example},~\ref{A_example},~\ref{Ac_example}
show the behavior of $G, G_c, A, A_c$ as functions of temperature
for different values of $V$ for the case of a Gaussian fields distribution
$P(h)=(2\Pi)^{-1/2}\exp(-h^2/2)$.

In case of a gap of amplitude $\Delta$ in the field distribution one
finds that both $A$ and $G$ vanish exponentially with that gap $G\sim
T\exp(-\beta \Delta)$ and the conjecture does not hold anymore.

We will now prove that, under some general conditions, the
conjectured zero-temperature values for $G$ and $G_c$ hold for any
disordered system.

\section{A proof of the conjecture}

To generally prove (\ref{eq7}) and (\ref{eq9b}) we start by considering
a general Hamiltonian ${\cal H}(\lbrace\s\rbrace)$ where the
$\lbrace\sigma_i;i=1,..,V\rbrace$ are Ising variables which may take the
values $\pm 1$ \cite{FOOTNOTE1}. This Hamiltonian may be written in
terms of the local fields,

\be
{\cal H}=-\sum_{i=1}^V\,h_i\s_i~~~,
\label{eq18b}
\ee

\noindent
where the $h_i$ are local fields proportional to

\be
h_i\propto \frac{\partial {\cal H}}{\partial \s_i}~~~,
\label{eq19b}
\ee

\noindent
which depend on the configuration $\lbrace\s\rbrace$. Suppose now that
the Hamiltonian ${\cal H}$ may only take continuous values so there is
no ground state degeneracy (apart from a global symmetry in the
Hamiltonian such as time-reversal symmetry; this case will be
discussed below). In particular, no local field $h_i$ vanishes. Let us
denote by $\lbrace\s^*\rbrace$ the (unique) ground state
configuration. The ground state is stable with respect all possible
number of spin flips, so that the value of the energy in that
configuration ${\cal H}(\lbrace\s^*\rbrace)$ is an absolute
minimum. In particular the ground state is stable respect to single
spin flips and the local fields evaluated at the ground state
satisfy the property,

\be
\s_i^*=sign(h_i^*) 
\label{eq20b}
\ee

\noindent
where $h_i^*$ are evaluated at $\s^*$.  To prove
the conjecture we need to prove the following statement:

\begin{itemize}

%\item{Statement 1.} The local fields (\ref{eq19}) evaluated in the ground
%state configuration $\lbrace\s^*\rbrace$ are uncorrelated from sample to
%ample at different sites.

\item {\bf Statement:} Excitations which involve the reversal of a single spin
yield the dominant contribution to the low-temperature behavior for all the quantities 
$(\overline{q^k})^l$ for any positive integers $k,l$ and by extension, to
the numerator and denominator in (\ref{eq5},\ref{eq6},\ref{eq8},\ref{eq9}).

\end{itemize}

%The first statement is consequence of the disordered character of
%the ground state. Mathematically speaking it means the following,

%\be
%\overline{(h_i^*)^m(h_j^*)^n}=\overline{(h_i^*)^m}\;
%\overline{(h_j^*)^n}~~~~~~i\ne j
%\label{eq21}
%\ee

%for any pair of positive integers $m,n$. Automatically this implies that
%the local field distribution is random and uncorrelated much alike in
%the instructive example of previous section (\ref{eq10}).

This statement somehow allows to map the most probable excitations in
(\ref{eq18b}) with those of the instructive example presented before
(\ref{eq10}). Nevertheless, we must emphasize two points. The first one
is that the ground-state local field distribution in the previous example
(\ref{eq10}) was taken uncorrelated for different sites and also the
same distribution was taken for each spin $i$. In general this is not
true. Local fields at different sites may be correlated and the
distribution on a given site may depend on the site. For instance, in
models with open boundaries the local field distribution for the sites
located on the surface is certainly different from the distribution of
those in the bulk.  The second observation is that, in general, the
lowest excitations in (\ref{eq18b}) may involve groups of several spins
(and not a single spin flip like in the simple case (\ref{eq10})).  So in
order to prove the conjecture we must show that excitations in
(\ref{eq18b}) which involve the reversal of any number of spins larger
than one always yield subdominant low-temperature corrections to the
single-spin excitation case.

%Once this result is generally proved the mapping from the general
%problem (\ref{eq18}) to the simple case (\ref{eq10}) is possible and the
%{\em universal} quantities (\ref{eq7},\ref{eq9b}) are recovered.

In what follows we present a constructive
proof of the previous statement without need to refer to the results of the
instructive example which had some restrictive assumptions. We start from the
general Hamiltonian (\ref{eq18b}) and analyze the low-temperature
behavior of the order parameters $\overline{\langle
q^2\rangle},\overline{\langle q^4\rangle}$ and $\overline{\langle
q^2\rangle^2}$. We will first consider the case of one spin
excitations and later on the more general one of higher-order
excitations. 

\begin{itemize}

\item{\em One-spin excitations}

The calculation proceeds as follows. Consider the ground state $\lbrace
\s^* \rbrace$ of (\ref{eq18b}) as {\em unique} and one-spin excitations
which involves the reversal of a single spin. If we consider the ground
state plus this class of $V$ possible excitations we can compute the
correlation function $\langle\s_i\s_j\rangle$ ($i\ne j$), obtaining the
result:

\be
\langle\s_i\s_j\rangle=\s_i^*\s_j^*\Bigl (1-2\frac{\exp(-2\beta h_i^*\s_i^*)
+\exp(-2\beta h_j^*\s_j^*)}{1+\sum_{l=1}^V\exp(-2\beta h_l^*\s_l^*)}\Bigr )
\label{eq22}
\ee

Defining $x_i=\exp(-2\beta h_i^*\s_i^*)$, we get in the $\beta\to\infty$
limit,

\be
\langle\s_i\s_j\rangle=\s_i^*\s_j^*(1-2(x_i+x_j))~~~~~(i\ne j),
\label{eq23}
\ee

where we have aproximated by 1 the term in the denominator of the
ratio in (\ref{eq22}). Such an approximation is allowed provided one
performs the limit $\beta\to\infty$ {\it before} the infinite volume
limit. Note that, in that denominator, each exponential contributes to
the sum at most with a term proportional to the temperature (see
below). Because there are $V$ terms of that type, at most that term is
of order $VT$. Hence, in the limit $TV\ll 1$, that denominator equals 1.
The result (\ref{eq23}) is the only quantity we need in order to
evaluate $\langle q^2\rangle$ and $\langle q^2\rangle^2$. In terms of
the variable $T_{ij}=\langle \s_i\s_j\rangle^2$, these are given by,

\be
\langle q^2\rangle = \frac{1}{V}+\frac{1}{V^2}\sum_{i\ne
j}T_{ij}~~~~,
\label{eq24}
\ee

\be
\langle q^2\rangle^2=\frac{1}{V^2}+\frac{2}{V^3}\sum_{i\ne j}T_{ij}
+\frac{2}{V^4}\sum_{i\ne j}T_{ij}^2+\frac{4}{V^4}\sum_{(i\ne j \ne
k)}T_{ij}T_{ik} +\frac{1}{V^4}\sum_{(i\ne j\ne k\ne l)}T_{ij}T_{kl}~~~~,
\label{eq25}
\ee

where the indexes in the sums run from $1$ to $V$ and correspond to
different sites.  To average (\ref{eq24}),(\ref{eq25}) over the
disorder we need to compute disorder averages of terms of the type
$x_i^mx_j^n$ where $i,j$ denote sites and $m,n$ positive integers. It
is easy to show that, in the absence of gap in the ground-state
local-field distribution, the terms with $i=j$ yield the dominant
low-temperature corrections and vanish linearly with $T$. Terms with
$i\ne j$ yield higher-order $O(T^2)$ contributions.  Suppose
$P(\lbrace h_i^*\rbrace)$ stands for the ground-state local-field
probability distribution.  For the terms $x_i^mx_j^n$ ($i\ne j$), we
have

\be
\overline{x_i^mx_j^n}=\int_{-\infty}^{\infty}
\exp(-2m \beta h_i^*\s_i^*)\exp(-2n\beta
h_j^*\s_j^*)P(h_1^*,..,h_V^*)dh_1^*..dh_V^*~~~~~.\label{eqa}
\ee

The field variables $h_k^*~~(k\ne i,j)$ may be integrated out, yielding
the following expression

\be \overline{x_i^mx_j^n}=\int_{-\infty}^{\infty} \exp(-2m \beta
h_i^*\s_i^*)\exp(-2n\beta
h_j^*\s_j^*)\,\hat{P}_{ij}(h_i^*,h_j^*)dh_i^*dh_j^*~~~~,\label{eqb0}\ee

\be
\hat{P}_{ij}(h_i^*,h_j^*)=
\int_{-\infty}^{\infty}\, P(h_1^*,..,h_V^*)\prod_{k\ne (i,j)}dh_k^*~~~~~~.
\label{eqb}
\ee

%Define the reduced ground state local field distribution as the
%single-site distribution,

If the local field distribution $P(\lbrace h_i^*\rbrace)$ has finite
weight at the point $h_i=0\,\forall i$, then the same holds for the two-sites
probability $\hat{P}_{ij}(0,0)$ so that we may expand this term around
$(0,0)$ in (\ref{eqb0}) obtaining thereby,

\be
\overline{x_i^mx_j^n}=\int_{-\infty}^{\infty}
\exp(-2m \beta h_i^*\s_i^*)\exp(-2n\beta
h_j^*\s_j^*)\bigl (\hat{P}_{ij}(0,0)+\Bigl (\frac{\partial
\hat{P}_{ij}}{\partial h_i^*}\Bigr)_{(0,0)}h_i^*+\Bigl( \frac{\partial
\hat{P}_{ij}}{\partial h_j^*}\Bigr)_{(0,0)}h_j^*+O(h_i^*h_j^*)\bigr)\,dh_i^*dh_j^*~~~~~~,
\label{eqd}
\ee

where $O(h_i^*h_j^*)$ denote higher-order terms at least quadratic in the
fields. A simple saddle-point
calculation (in the $\beta\to\infty$ limit) gives then,

\be
\overline{x_i^mx_j^n}=\frac{T^2}{mn}\hat{P}_{ij}(0,0)+O(T^3)~~~.
\label{eqe}
\ee

The dominant terms in the limit $T\to 0$ 
correspond to terms of the type $\overline{x_i^n}$, which give

\be
\overline{x_i^n}=\frac{T\hat{P}_i(0)}{n}~~~~,
\label{eq26}
\ee

where $\hat{P}_i(0)$ is the value of the single-site probability
distribution on the site $i$ evaluated at $h=0$,

\be
\hat{P}_i(h^*)=\int_{-\infty}^{\infty}P(h_1,..,h_V)\delta(h_i-h^*)
\prod_{k\ne i}dh_k~~~.
\label{eqc}
\ee

%\be
%P_i(h)=\overline{\delta(h-h_i^*)}
%\label{eq27}
%\ee

This probability is not independent of the spin $i$, as our Hamiltonian
can contain terms which introduce asymmetry between different
sites. This is an important difference with respect to the computation of the
previous section where the local field distribution (\ref{eq11}) was
site independent.  Actually, this independency was necessary in the
``instructive example'' to fully carry out the analytic computation of $G$
and $G_c$.  The key point is that, at the level of one-spin excitations,
low-temperature corrections to overlap averages are linear in $T$ and
$\hat{P}_i(0)$. According to expressions (\ref{eq24}), (\ref{eq25}) all sites
are equivalent (inequivalence of sites enters only through the value of
$\hat{P}_i(0)$), so the only invariant term linear in $P$ is
$\sum_i\hat{P}_i(0)$.  The numerator in (\ref{eq5}) yields,

\be
\overline{\langle q^2\rangle^2}-\overline{\langle q^2\rangle}^2 = 
\frac{16 T \sum_i P_i(0)}{3V^4}(V-1)^2\,+\,O(T^2)~~~~~~.
\label{eq28}
\ee

To compute the overlap 
$\overline{\langle q^4\rangle}$ we use the expression,

\be
\overline{\langle q^4\rangle}=\frac{1}{V^4}\Bigl ( 3V^2-2 V +
(6V-8)\sum_{i\ne j}T_{ij} +\sum_{(i,j,k,l)}T_{ijkl}\Bigr )~~~~,
\label{eq29}
\ee

where $T_{ijkl}=\langle \s_i\s_j\s_k\s_l\rangle^2$. Similarly as for the
two-point correlation function (\ref{eq23}) we obtain

\be
\langle\s_i\s_j\s_k\s_l\rangle=\s_i^*\s_j^*\s_k^*\s_l^*(1-2(x_i+x_j+x_k+x_l))~~~~~(i,j,k,l~~$all~~different$)~~~.
\label{eq30}
\ee

\noindent
With the same assumptions as for the two-points function we
obtain for the denominator in (\ref{eq4})

\be
\overline{\langle q^4\rangle}-\overline{\langle q^2\rangle}^2 = 
\frac{16 T \sum~_i P_i(0)}{V^4}(V-1)^2\,+\,O(T^2)~~~~~,
\label{eq31}
\ee

which finally yields,

\be
G=\frac{1}{3}\,+\,O(T)~~~~~.
\label{eq32}
\ee

A similar calculation for $G_c$ yields $G_c=\frac{13}{31}+O(T)$.

\item{\em Two-spin excitations}

Let us consider now excitations which involve  only
two different spins in the lattice ($V(V-1)/2$ different type of
excitations).  In this calculation one-spin excitations are not
included. It is easy to check that these excitations yield $O(T^2)$
corrections to the two-spin and four-spin correlations. Under the same
conditions as before these are given by,

\be
\langle\s_i\s_j\rangle=\s_i^*\s_j^*(1-4(x_i+x_j)\sum_{l\ne i,j}x_l)~~~~~(i\ne j)
\label{eq33}
\ee

\be
\langle\s_i\s_j\s_k\s_l\rangle=\s_i^*\s_j^*\s_k^*\s_l^*(1-4(x_i+x_j+x_k+x_l)\sum_{m\ne
i,j,k,l}x_m)~~~~~({\rm i,j,k,l~~~all~~different})
\label{eq34}
\ee

A saddle point calculation shows that corrections to the ground-state
correlation functions are quadratic in $T$. Finite $T$ corrections now
depend on both $\overline{x_i}\,\,\overline{x_j}$ and $\overline{x_ix_j}$
for $i\ne j$. Now, for the quantity $G$ we expect a dependence of both
numerator and denominator on terms of
the type $\hat{P}_i(0)\hat{P}_j(0)$ as well as $\hat{P}_{ij}(0,0)$. They
can enter in different forms, for instance $\sum_{i\ne
j}\hat{P}_{ij}(0,0)$, $(\sum_i\hat{P}_i(0))^2$ or
$(\sum_i\hat{P}_i(0)^2)$. A universal value for $G$ is not guaranteed
anymore. In particular, supposing uncorrelated local fields (which in principle
may not be true) and independency of the one-site probability
distribution $\hat{P}_i(0)$ on the site $i$ we obtain, after a simple but lengthy
calculation,

\be
\overline{\langle q^2\rangle}^2-\overline{\langle q^2\rangle}^2 = 
\frac{128 T^2 P(0)^2}{9V^3}(V-2)^2(V-1)\,+\,O(T^3)
\label{eq35}
\ee

\be
\overline{\langle q^4\rangle}-\overline{\langle q^2\rangle}^2 = 
\frac{64 T^2 P(0)^2}{V^3}(V-2)^2(V-1)\,+\,O(T^3)
\label{eq36}
\ee

and their ratio yields $G=\frac{2}{9}\,+\,O(T)$ which is different than
before. We stress again that the result $2/9$ is not universal and will
certainly not hold in the most general case. This calculation has been
shown to stress how the $1/3$ value is a fingerprint of the dominancy in
the limit $T\to 0$ of the one-spin excitations.

\item{\em Higher-order excitations}

The generalisation to the most general case of $K$-spin excitations is
straightforward.  Including only this class of excitations we obtain
$O(T^K)$ corrections to correlations which involve any finite number of
spins. This can be easily seen from the fact that any possible
excitation of this type will involve the reversal of $K$ different
spins, each spin $i$ contributing by a factor $x_i=\exp(-2\beta h_i^*)$
to the correction. The simultaneous effect of all spins yields a product
type $\prod_{i=1}^K x_i$ which immediately gives (in the limit
$\beta\to\infty$) the $T^K$ term. The numerator and denominator in $G$
are of order $T^K$ with $O(T^{K+1})$ corrections. The final result for $G$
for any value of $K$ is not easy to compute and, as previously
discussed, will depend on a larger number of invariants which involve
different combinations of the terms 
$\hat{P}_i(0),\hat{P}_{i_1i_2}(0,0),..,\hat{P}_{i_1,..i_K}(0,0)$.

\end{itemize}

When all possible excitations are treated together the calculation
proceeds as before. The dominant contribution for OPF will always come
from samples whose lowest excitations are one-spin
excitations. Consequently, in the zero-temperature limit (for $V$
finite) one-spin excitations dominate the correction to correlation
functions proving our conjecture. Note that the result we are stating
here is quite natural. OPF at very low temperatures are always dominated by those rare
samples characterised by local fields $\beta h <<1$ 
where one spin-excitations yield the largest contribution. From
a numerical point of view this implies that more samples are needed to
compute with a reasonable precision the values of $G$ and $G_c$ as $T$
goes down. This is because for $T\to 0$ the effect from rare samples on
OPF becomes more and more important. Let us stress again that the
present derivation assumed that $TV\ll 1$. In the opposite limit or in an
intermediate regime the result obviously does not hold. In that case, it
may well happen that dominant contributions in OPF involve the reversal
of a large number of spins (domain excitations) which, in the limit
$TV\gg 1$, may also involve the whole system \cite{KM}.

The hypothesis of a unique-ground state is apparently in contradiction with
the case in which there is time-reversal symmetry. Indeed
all spin-correlations computed in this section are invariant under
time-reversal symmetry and the present conclusions remain unchanged. The
situation is certainly different in disordered systems with
non-trivially degenerate ground states (for instance, finite-dimensional
spin glasses with discrete couplings) where we expect that $G(V,T)$
vanishes exponentially with $1/T$ like in the instructive example of the
previous section. Again, in the other limit (finite temperature and
$V\to\infty$) the behavior of these degenerate models may completely
change and $G$ could be finite again \cite{FOOTNOTE2}.

\section{The 1D Ising spin glass}

In this section we present an analysis of the 1D-Ising spin-glass 
model with free boundary conditions. We consider the folllowing hamiltonian:
\be
{\cal H}=-\sum_{i=1}^{N-1}J_{i}\sigma_{i}\sigma_{i+1}~~~~,
\ee   

where the couplings are randomly distributed according to the
probability distribution $P(J)$.  Our aim is to obtain an analytic
expression for $G$ and $A$ eqs. (\ref{eq5}) and (\ref{eq6}). As this model has the
transition at $T=0$, we expect that in the large volume limit $G$ will
go to zero except at $T=0$, where $G=1/3$. Moreover, we show that at
zero temperature $G=1/3$ for any finite system, although here the two
limits ($V\to\infty$ and $T\to 0$) do not commute. In order to obtain
an expression for the moment of the order parameter $q$ we have computed
the following object:

\be \overline{{\langle e^{yq}\rangle }^m}~~~~~~,
\label{eyq}
\ee

where $m$ is a positive integer and $q$ is the overlap between two
different configurations of spins, which is the generator of the moments
of the overlap $\overline{{\langle q^{p}\rangle}^s}$. Once obtained this
expression, by partial derivation respect to $y$ we will obtain
expressions for the expectation values of all the moments of $q$, such as:

\be
\overline{\langle q^n \rangle}={\frac{\partial ^{n}\overline{{\langle e^{yq}\rangle}}}{\partial y ^{n}}}|_{y=0}~~~~~.
\label{mom}
\ee In our computation we are only interested on the quantities:
$\overline{\langle q^2\rangle},\overline{\langle q^4\rangle}$ and
$\overline{\langle q^2\rangle^2}$. Consequently we only need to compute
(\ref{eyq}) for $m=1,2$.  The former can be easily computed by
(\ref{mom}). By doing some more work we can obtain an expression for
$\overline{\langle q^{2}\rangle^{2}}$:

\be \overline{\langle
q^{2}\rangle^2}=\frac{1}{3}\left[\frac{\partial^4\overline{\langle
e^{yq}\rangle}^2}{\partial y ^{4}}-\frac{\partial
^{4}\overline{\langle e^{yq}\rangle}}{\partial y
^4}\right]_{y=0}~~~, \ee

where we have used the fact that in this model $\langle q\rangle\;=\;0$.

\subsection{The transfer matrix method}
For general $m$, (\ref{eyq}) can be computed through the transference matrix method \cite{BM1}. 
We have to compute:
\be
\prod_{\alpha=1}^{m}\frac{\sum_{\{\sigma^{\alpha}\} \{\tau^{\alpha}\}}\exp\left(y\sum_{i=1,N}\frac{\sigma_{i}^{\alpha}\tau_{i}^{\alpha}}{N}+ \beta \sum_{i=1}^{N-1}{J_{i}(\
	\sigma_{i}^{\alpha}\sigma_{i+1}^{\alpha}+\tau_{i}^{\alpha}\tau_{i+1}^{\alpha}})\right) }{{\cal Z}^{2}}
\ee
where  ${\cal Z}$$ = 2 \prod_{i} 2 \cosh(\beta J_{i})$ is the partition
function of a 1D chain, $\alpha$ is the index for each pair of
replicas and we have $m$ systems of two replicas.\\

In order to perform the average over the disorder, we are interested in considering the transfer matrix associated to each point $i$, so that it contains all the dependence of the $J_{i}$. For a single pair of replicas this matrix reads:

\be
V_{i}\equiv V(\sigma_{i},\tau_{i};\sigma_{i+1},\tau_{i+1})= 
\frac{exp\left(y\frac{\sigma_{i}\tau_{i}+\sigma_{i+1}\tau_{i+1}}{2 N}+ \beta J_{i}(\
	\sigma_{i}\sigma_{i+1}+\tau_{i}\tau_{i+1})\right)}{(2 \cosh(\beta J_{i}))^{2}}~~~~~.
\label{t}
\ee 
For general $m$ our matrix associated to each point consists of the
tensorial product of $m$ matrices $V_{i}$. 
At this stage we are ready to perform the average over the disorder and for any $i$ we have:

\be 
\overline{T}= \overline{T_{i}}= \overline{\bigotimes_{1}^{m} V_{i}}
\ee 
Then our calculation is reduced to:
\be
\overline{{\langle e^{yq}\rangle}^m}=\frac{1}{4}\sum e^{y \frac{\sum_{\alpha}{\sigma_{1}^{\alpha}\tau_{1}^{\alpha}}}{2 N}}{\overline{T} }^{N-1}e^{y \frac{\sum_{\alpha}{\sigma_{N}^{\alpha}\tau_{N}^{\alpha}}}{2 N}}~~~~,
\ee
so we must compute the trace of the product
\be
{\overline{T} }^{N-1} B~~~~~,
\ee 
where $A$ is a $ 4m\times 4m$ matrix, which is the tensorial product of $m$ matrices, which contain the terms of the two edges which had fallen out in the symmetrization process,  
\be
B=\bigotimes_{\alpha}\frac{1}{2^{2 }}e^{y \frac{\sigma_{1}^{\alpha}\tau_{1}^{\alpha}+\sigma_{N}^{\alpha}\tau_{N}^{\alpha}}{2 N}}.
\ee 
The rest of the calculation is straightforward: In first place we have
to diagonalise $\overline{T}$, and obtain the set of eigenvalues and
eigenvectors, so that in this new base we have

\be
{\overline{T}_{\lambda}}^{N-1}=\left(
\begin{array}{ccc}
\lambda_{1}^{N-1} & ......&  .......\\ 
..... & \lambda_{2}^{N-1} & ....... \\
..... & ........ & ....... \\
..... & ...... &  \lambda_{2^{2m}}^{{\tiny N-1}}
\end{array}
\right)~~~~~,
\ee

\noindent
where the subindex $\lambda$ stands for the diagonalised matrix.
We then have to obtain the change of base matrix $M$ which expresses the new set of eigenvectors $\lbrace{\lambda^{i}}\rbrace$ in terms of the old base $\lbrace{\s^{\alpha}}\rbrace$.
We finally obtain: 
\be
\overline{{\langle e^{yq}\rangle }^m}=Tr \;M {\overline{T}_{\lambda}}^{N-1}M^{T}B .
\ee  

We have to point out that the case $m=1$ is easy to solve. However, the case $m=2$ turns up to be more difficult as the diagonalization of $\overline{ V}$ is not trivial. To compute $\overline{\langle q^2\rangle^2}$ one can always use the traditional method by using the fact that
\be
\langle \s_i\s_j \rangle\;=\;\prod_{p=i,j-1} \tanh \beta J_p \;\;\; i\neq j~~~.
\ee
\subsection{Results}

Here we report on the obtained results in the
low-temperature limit and in the infinite-volume limit.  The relevant 
quantities $\overline{\langle q^2\rangle},\overline{\langle
q^4\rangle}$ and $\overline{\langle q^2\rangle^2}$ only depend on $N$,
$\overline{R}$ and $\overline{R^2}$ which have been introduced in
section III, and whose low-temperature behavior is given by
(\ref{eq16}).  At finite temperature, where $\overline{R}^{N}$ and
$\overline{R^{2}}^{N}$ $\ll 1$ we obtain for the numerator and denominator
in (\ref{eq5}):

\be
numerator=\frac{4(1+\overline{R})(\overline{R}^2- \overline{R^{2}})}{N^{3}(\overline{R}-1)^{3}(\overline{R^{2}}-1)}+ {\cal O}(\frac{1}{N^{2}})~~~~~,
\ee     
\be
denominator=\frac{4(1+\overline{R})(\overline{R}^2-\overline{R^{2}})}
{N^{2}(\overline{R}-1)^{3}(\overline{R^{2}}-1)}+ {\cal O}(\frac{1}{N^{3}})~~~~,
\ee     

\noindent
where we have kept the lowest orders in $1/N$ and we have made the
following aproximations $\lim_{N \rightarrow \infty}
\overline{R}^{N},\overline{R^{2}}^{N} \rightarrow 0$. We see that in
this limit $G$ goes to zero as $1/N$.  However, if we take the low
temperature limit (\ref{eq16}), where $\overline{A},\overline{A^{2}} \approx 1$
then we get the expressions

\be
numerator=\frac{4 D (N^{4}-1)T}{45 N^{3}}+ {\cal O}(T^{2})~~~~,
\ee     
\be
denominator=\frac{4 D(N^{4}-1)T}{15 N^{3}} + {\cal O}(T^{2})~~~~,
\ee 
    
\noindent
where $D$ is given by $D=2P(0)$. This yields $G=\frac{1}{3} + {\cal
O}(T)$, independently of the size of the system. A detailed
computation up to second order in $T$ gives us that in the large
volume limit: $G=\frac{1}{3} - B TN $, $B$ being a constant. In fact
for the parameter $A$, we get in the limit $T \rightarrow 0$:

\be
A=\frac{4 D (N^{4}-1)T}{45 N^{3}}+ {\cal O}(T^{2})~~~~~~.\label{A_marta}
\ee     

\begin{figure}
\begin{center}
\rotatebox{270}
{\epsfxsize=7cm\epsffile{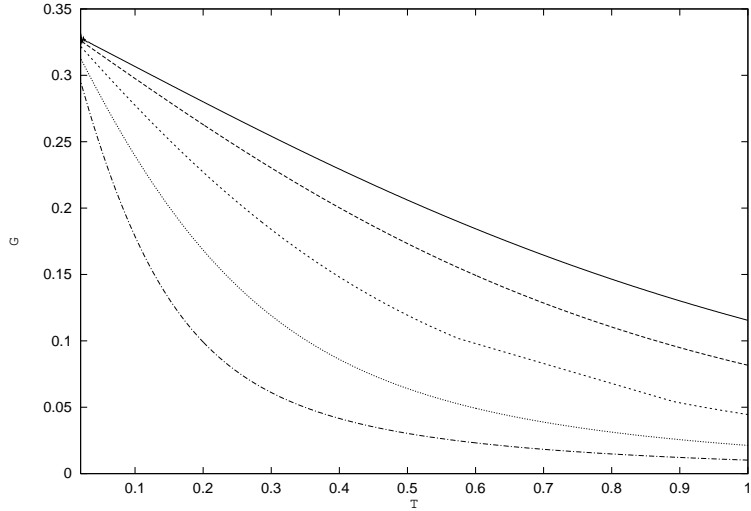}}
\caption{Parameter $G$  for the 1D Ising spin glass for lengths N=2,4,8,16,32 (from above to below)}
\label{1d_g}
\end{center}

\end{figure}

\begin{figure}
\begin{center}
\rotatebox{270}{\epsfxsize=7cm\epsffile{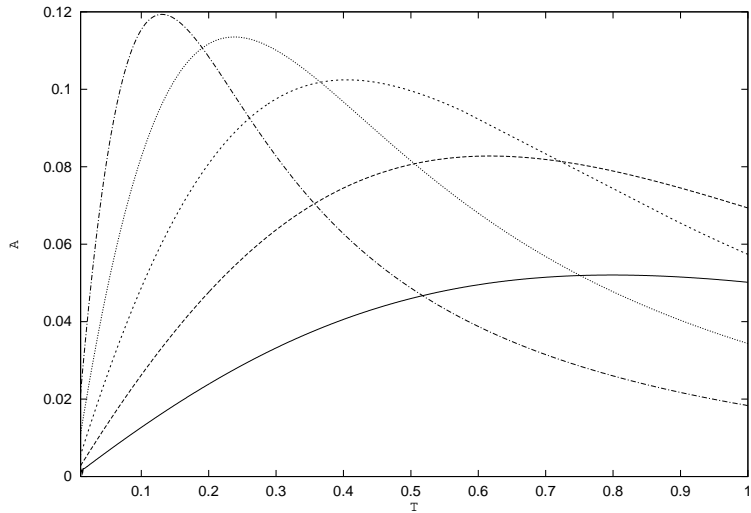}}
\caption{Parameter $A$  for the 1D Ising spin glass for lengths N=2,4,8,16,32 (from right to left)}
\label{1d_a}
\end{center}

\end{figure}

In figures~(\ref{1d_g}) and (\ref{1d_a}) we show $G$ and $A$ as a
function of the temperature for a Gaussian distribution odf couplings
$P(J)=(2\Pi)^{-1/2}exp(-J^2/2)$.  Note that the low-temperature
correction to $G$ and (\ref{A_marta}) scale as $TN$ when $N\rightarrow
\infty$ reflecting the fact that as we get close to the transition
point $T=0$, the correlation length diverges as $1/T$. We recover the
desired result at $T=0$, however we have to stress out that in this
model both limits $T\to 0$ and $N\to\infty$ do not commute.

\section{The spherical Sherrington-Kirkpatrick spin glass}

In this section we present some numerical simulations for the values
of $G$ and $A$ in the Sherrington-Kirkpatrick (SK) spherical spin
glass. This case is quite interesting because its low-temperature
behavior corresponds to the second possibility mentioned in section II
where OPF vanish (in the $V\to\infty$ limit) much slower than the
paramagnetic example studied in the previous section. Correspondingly
the study of OPF in this model turns out to be very complicated
because the equilibrium solution is marginally stable.  The model is
defined by \cite{KTJ}

\be
{\cal H}=-\sum_{i<j}\,J_{ij}\s_i\s_j~~~~~,
\label{eqs1}
\ee

\noindent
where $-\infty< \s_i< \infty$ and the values of $\s_i$ satisfy the
spherical global constraint $\sum_{i=1}^N\s_i^2=N$. The couplings have
average zero and variance $1/N$. The statics of this model can be solved
with and without replicas \cite{KTJ}. In the former case one finds a
transition at a temperature $T_c=1$ where the Edwards-Anderson parameter
is different from zero and equal to $1-T$. In the latter case the
transition corresponds to a macroscopic condensation of spin
configurations onto the eigenvector corresponding to the largest
eigenvalue. In the replica framework it has been shown \cite{BRAY} that
the replica symmetric solution is the only possible one within the
Parisi scheme. Since OPF vanish, the computation of $G$ requires the
knowledge of finite-size corrections in the numerator and denominator in
(\ref{eq5}). A simple calculation reveals that the replica symmetric
solution is marginally stable (the replicon eigenvalue vanishes
everywhere below $T_c$) so the spin-glass susceptibility diverges. The
situation is similar to what happens in the usual
Sherrington-Kirkpatrick model with Ising spins. There the spin-glass
susceptibility diverges proportionally to the volume while now the
divergence is much slower (like $N^{1/3}$). This is so because in the
present model OPF vanish like $N^{-2/3}$ while in the original SK model
OPF are finite.

Again, to compute $G$ we need to know the precise value of the
 amplitudes entering in the finite-size corrections in the parameters
 $\overline{\langle q^2\rangle}, \overline{\langle q^2\rangle^2},
 \overline{\langle q^4\rangle}$. It is well known that analytic
 calculations of finite-size corrections in spin glasses are extremely
 difficult, specially for the amplitudes which are the quantities we are
 interested in. For the SK model these amplitudes are partially known
 only for some quantities \cite{PRS}.  For the present case we will use
 theoretical considerations and numerical simulations to estimate the
 asymptotic behavior of the different overlaps.

We have simulated model (\ref{eqs1}) with a Monte Carlo dynamics where
 a change of a randomly chosen spin is proposed $\s_i\to\s_i+\delta
r_i$ where $\delta$ is a constant number typically of order 1 and $r_i$
is a random number uniformly distributed between $-1/2$ and $1/2$. The
value of $\delta$ is chosen to have a reasonable acceptance rate. The
value of all other spins is recalculated in order to satisfy the global
spherical constraint. Moves are accepted according to the Glauber
algorithm. Note that although we need to recalculate the value of all
spins (changing them by multiplying by a normalization constant) the change
in the energy can be simply calculated in a finite number of operations
independent of $N$ and simulations are as fast as with Ising spins. Our
investigation has focused on small sizes, which reveal how $G$ is a
powerful tool to investigate phase transitions. The number of samples
simulated are typically several thousands for very small sizes
($N=4,6,8,12,16$) and several hundreds for larger ones
($N=24,32,40,48,64$). Overlaps have been computed by collecting
statistics over a large time window (tipically of order $10^5$ Monte
Carlo steps for each sample). We have evaluated, $\overline{\langle
q^2\rangle}^2, \overline{\langle q^2\rangle^2}, \overline{\langle
q^4\rangle}$ for different sizes and temperatures.

\begin{figure}
\begin{center}
\rotatebox{270}{\epsfxsize=8cm\epsffile{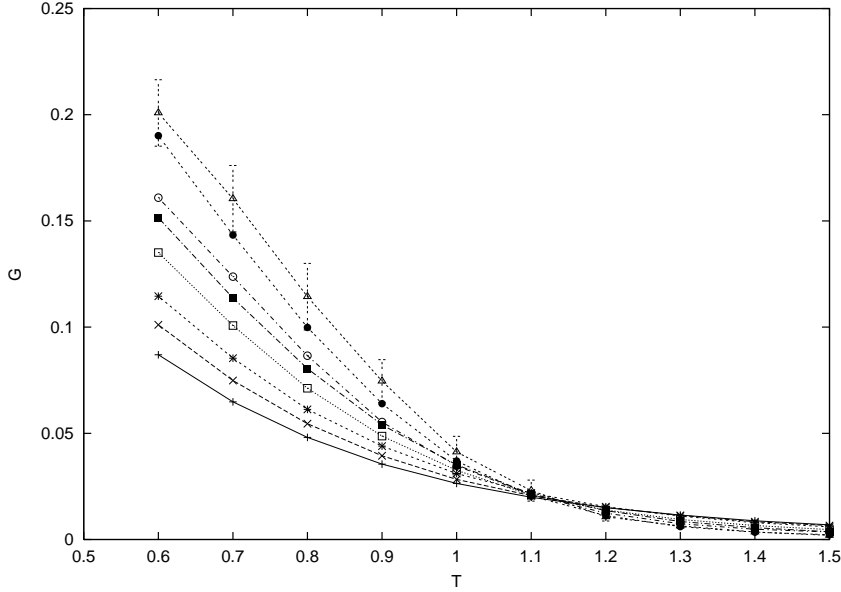}}
\caption{Parameter $G$  for the SK spherical spin glass ($N=4,6,8,16,24,32,48,64$ from below to above at low temperatures). The larges error bars are shown for the largest size $N=64$.}
\label{simG}
\end{center}

\end{figure}

\begin{figure}
\begin{center}
\rotatebox{270}{\epsfxsize=8cm\epsffile{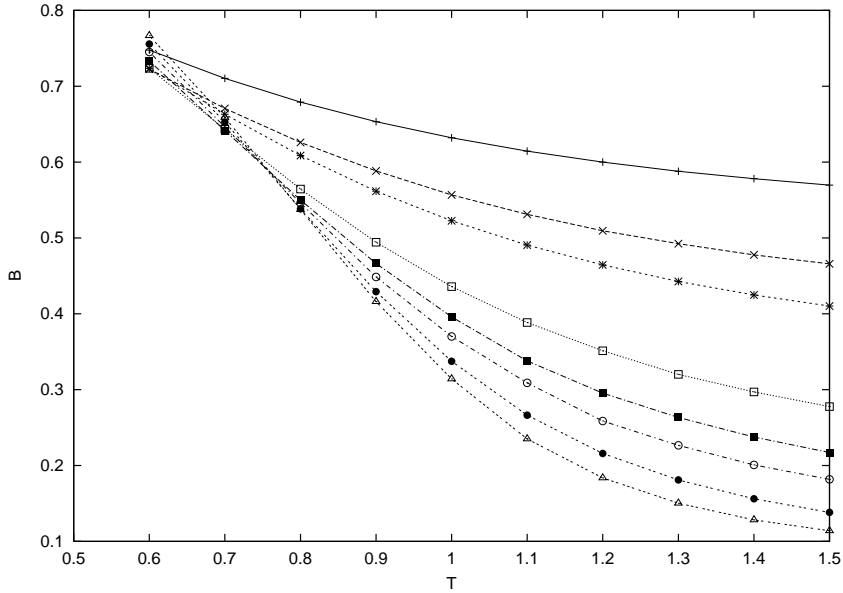}}
\caption{Binder cumulant $B$  for the SK spherical spin glass ($N=4,6,8,16,24,32,48,64$ from below to above at low temperatures). Error bars are now negligible.}\label{simB}
\end{center}

\end{figure}

Figure~\ref{simG} shows the results for $G$. Note that already for the
smallest sizes there is a crossing of the different curves. The crossing
appears for values of $T$ well above $T_c=1$ for the smallest sizes and
moves to lower temperatures as the size increases converging to the
expected value $T_c=1$. It is quite surprising that already for very
small sizes the transition can be clearly seen. The crossing moves down
in temperature as the sizes increase and already for several tens of
spins converges to the correct value $T=1$. As a comparison we show in
figure~\ref{simB} the behavior of the usual Binder parameter defined as

\be
B=\frac{1}{2}\Bigl (3-\frac{\overline{\langle q^4\rangle}}{\overline{\langle q^2\rangle}^2}\Bigr)
\label{eqB}
\ee

In this case the crossing point appears at low temperatures for small
sizes and moves up very slowly as the size increases. But already for
the largest sizes the crossing is still at $T\simeq 0.8$ quite far from
$T=1$. A similar effect has been observed in simulations of the
Sherrington-Kirkpatrick model with Ising spins \cite{BY,POTTS}. These results
indicate that a numerical study of the parameter $G$ can be extremely
useful to locate phase transitions in disordered systems by studying
very small sizes \cite{PRS2}.

To analyze better the behavior of $G$ at low temperatures we have tried
to extrapolate $G$ to the large $N$ limit. Below $T_c$ we expect for all
three quantities $\overline{\langle q^2\rangle}, \overline{\langle
q^2\rangle^2}, \overline{\langle q^4\rangle}$ the following finite-size
corrections,

\be
\overline{\langle q^2\rangle}^2, \overline{\langle q^2\rangle^2}, 
\overline{\langle q^4\rangle}= 
q_{EA}^4+\frac{a}{N^{2/3}}+\frac{b}{N}+\frac{c}{N^{4/3}}+
\frac{d}{N^{5/3}}~~~~,
\label{eqs2}
\ee

with $q_{EA}=1-T$.  From these expressions we expect for $G$ the
following behavior,

\be
G=G_{\infty}+\frac{A}{N^{1/3}}+\frac{B}{N^{2/3}}+O(1/N)~~~.
\label{eqs3}
\ee

\begin{figure}
\begin{center}
\rotatebox{270}{\epsfxsize=8cm\epsffile{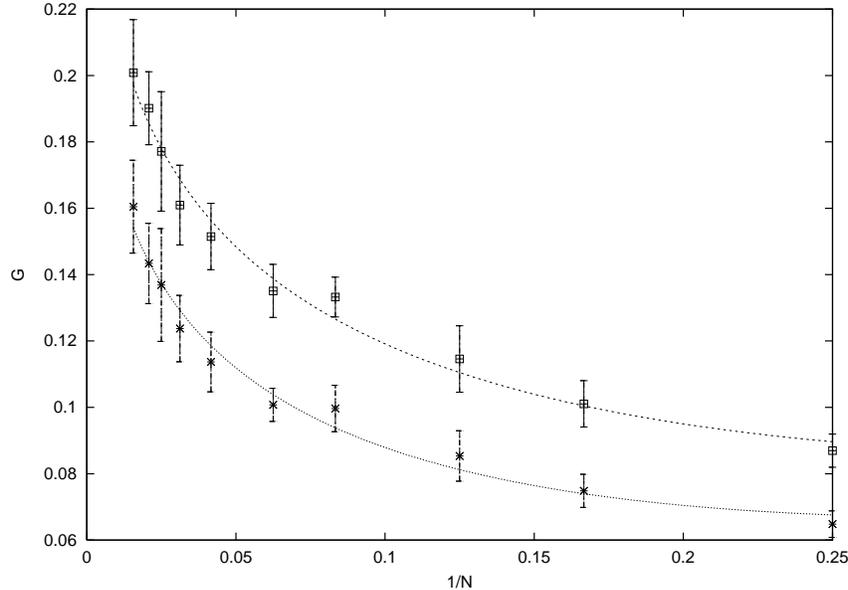}}
\caption{Fit function (\ref{eqs3}) to the $G$ parameter for different
sizes at $T=0.6$ (above) and $T=07$ (below). Extrapolations to
$N\to\infty$ are compatible with $G(V\to\infty)=1/3$ in the spin-glass
phase}\label{fits}
\end{center}

\end{figure}

We have fitted the values of $G$ to this expression with $G_{\infty}, A,
B$ as fitting parameters.  The results and the fits are shown in
figure~\ref{fits}. The extrapolated values for the lowest temperatures
$T=0.6, 0.7$ are $G_{\infty}(T=0.6)=0.34\pm 0.2,( A(T=0.6)=-0.71 \pm 0.1$
and $B(T=0.6)=0.49 \pm 0.13)$, $G_{\infty}(T=0.7)=0.29 \pm 0.2,(
A(T=0.7)=-0.66 \pm 0.1$ and $B(T=0.7)=0..49 \pm 0.12)$. Within errors
these are compatible with the value $1/3$. Trying to have an estimate of
$G_{\infty}$ at higher temperatures is very difficult because critical
effects are strong.

We must conclude that for this model the universal value $1/3$ is well
compatible with the data suggesting that this may be a generic result
for a spin-glass phase. Still we should do more extensive simulations to
reach a final conclusion. Although going to larger sizes at the lowest
temperatures may be factible this will require much longer computational
time.

\section{Outlook and discussion}

In this paper we have investigated order parameter fluctuations (OPF) in
spin glasses. In particular we have considered four different
parameters: $G,A$ for disconnected thermal averages and $G_c,A_c$ for
connected thermal averages. It has been recently shown that these models
can be very useful to investigate phase transitions in disordered
systems \cite{MNPPRZ,PPR} and several recent numerical works
\cite{BBDM,HK,POTTS,PPRICCI} indeed support this conclusion. In this work we
have concentrated our attention to obtain general results and to apply
them to certain solvable cases where these can be explictly checked.

We have demonstrated that for models with a unique ground state and no
gap in the ground-state local field distribution (for instance, all
discrete models with continuous disordered couplings taken from a
distribution without gap) $G$ and $G_c$ take the respective universal
values $G=1/3$, $G_c=13/31$ at zero temperature for any finite
volume. This is consequence of the dominancy of one-spin excitations
in OPF. For infinite volume this result still holds only in the regime
where the limit $T\to 0$ is taken before the limit $V\to\infty$ and
fast enough such that $TV\to 0$. This result has then been checked
calculating OPF in an instructive example without many body
interactions and for the case of the one-dimensional Ising spin glass
where explicit computations can be done using the transfer matrix
method. All these {\em good} properties suggest that both parameters
$G,G_c$ are ideal candidates to investigate phase transitions in
disordered systems much alike the Binder cumulant is for ordered
systems.

The extension of this result to the other limit where $V\to\infty$ is
taken before $T\to 0$ or, more generally, the limit $V\to\infty$ for
$T$ finite is far from trivial. In this last case, $G(V,T)$ is not
volume independent anymore. So the question is whether $G(V,T)$
converges in the large $V$ limit to the universal temperature
independent value 1/3.  At finite temperatures there are different
possible scenarios for the value of $G$. In case OPF are finite in the
$V\to\infty$ limit stochastic stability arguments and replica
equivalence suggest that $G$ should be 1/3 everywhere in the
spin-glass phase. Replica equivalence is a very generic property
which, to our knowledge, has not been emphasized before in the present
context and implies that the free energy of a replicated disordered
system must be proportional to the number of replicas.  Note that at
zero temperature replica equivalence cannot be used because the limits
$V\to\infty$ and $T\to 0$ may not commute in that case. Actually, as
we proved in section IV only for models with a unique ground state and
absence of gap in the fields distribution, $G$ takes the universal
value $1/3$ but vanishes (exponentially fast with $1/T$) in the
presence of a finite gap in that distribution.

The other interesting case is when OPF vanish. And here we can offer
only more speculative arguments. A possible scenario is that which
distinguishes two possibilities depending whether, in the
infinite-volume limit, OPF vanish like $1/V$ or slower like
$1/V^{\alpha}$ with $\alpha<1$. If OPF vanish like $1/V$, $G$ may take
the value 0 typical of a paramagnetic phase (for instance the case of
the one-dimensional spin-glass model) or a temperature dependent value
(the instructive example of section III). For these two solvable cases
the parameter $G$ is quite different. In the one-dimensional Ising spin
glass we find $G=\frac{1}{3}\delta_{T,0}$ while in the instructive
example we find $G=\hat{G}(T)$ with $\hat{G}$ a monotonous decreasing
function of $T$ with $\hat{G}(0)=1/3$. The reason for these two
different behaviors in a disordered phase may be adscribed to the fact
that, in the first case, there is a critical point at $T=0$ while in the
second there is no critical point at all. So $G$ is a good indicator for
a phase transition. But this observation must be taken with caution
because the parameter $G_c$ shows a different behavior for the
instructive example $G_c=\frac{13}{31}\delta_{T,0}$ similar to the
behavior of $G$ in the one-dimensional spin glass.  We expect the
interesting behavior to be present in models where OPF vanish like
$1/V^{\alpha}$ with $\alpha<1$. This class of models includes disordered
systems where the replica symmetric solution is marginally stable and
eventually finite-dimensional spin glasses if replica symmetry is not
broken, a question still unsolved\cite{REVIEW}. This case is much
more subtle because replica equivalence cannot be used (nor probably the
stochastic stability property) and finite-size corrections must be
known. To address this question we have done a numerical study of the
spherical Sherrington-Kirkpatrick spin glass. Two are the main outcomes:
1) The parameter $G$ is an excellent tool to locate the spin-glass
transition already for very small sizes (more precise than the usual
Binder parameter) and 2) An infinite-volume numerical extrapolation
(compatible with the expected form for the finite-size corrections) of
the value of $G$ in the spin-glass phase is well compatible with the
value 1/3.

shares the same p

Before concluding we want to stress that, apart from their applicability
to the study of spin-glass transitions, OPF are interesting quantities
which deserve further investigation. The outcome of the proof in section
IV is that OPF are much sensitive and rely completely on the effect of
rare samples. Actually, rare samples are those which induce the
largest OPF and fix the value of $G$ to $1/3$. A comprehensive study of rare
events in disordered systems is still missing. Averaging of extensive
quantities such as the replicated free energy in standard
renormalization group approaches may wipe out a large number of effects such
as those discussed here.  Certainly more detailed investigations are
needed to clarify the situation.  Although a final theorem which
resolves this problem may be at hand we think that the search for
non-trivial counterexamples of the different possibilities discussed in
this paper could be very useful.

{\bf Acknowledgments}.  We acknowledge discussions with M. Picco and
A. J. Bray. We are indebted to A. A. Garriga and A. Rocco for a careful reading of the
manuscript. F.R is supported by the Ministerio de Educaci\'on y Ciencia
in Spain (PB97-0971). M. S. is supported by the Ministerio de
Educaci\'on y Ciencia of Spain, grant AP-98 36523875.

\hspace{-2cm}

%\end{multicols}
\end{document}